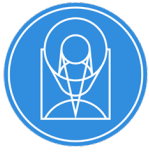

# JWST TECHNICAL REPORT

| Title: Effects of Jump Detection and Ramp Fitting Algorithms on NIRISS/SOSS Exoplanet Time-Series Observations | | Doc #: JWST-STScI-008975, SM-12<br>Date: 26 March 2025<br>Rev: - |
|---|---|---|
| Authors: Aarynn Carter, Néstor Espinoza, Loïc Albert (Université de Montréal), Tim Brandt, Tyler Baines, Joe Filippazzo, Kevin Volk | Phone: (410) - 338-4758 | Release Date: 6 May 2025 |

## 1 Abstract

Jump detection and ramp fitting are fundamental steps when elevating JWST data products from up-the-ramp measurements to integration level count rate images. Occurring at an early stage of the overall data reduction and calibration framework, these steps, and any biases introduced by them, have the potential to significantly impact end-stage scientific measurements. Here we explore the differential impacts of jump detection and ramp fitting for the current default JWST pipeline implementation versus a newly developed, likelihood-based method, in the context of NIRISS/SOSS exoplanet time-series observations. Across both on-sky data and a suite of simulated data, we find that the two methods show differences in end product transmission spectra of order ~10's to ~100's parts-per-million. For the simulated data in particular, we show that the likelihood method offers a 12-18% improvement in the residual scatter and maximum deviations from the simulated ground truth. Finally, by repeating our simulations across various noise prescriptions, we corroborate the need for group-level $1/f$ noise corrections and re-emphasize the default implementation of such procedures as a key priority for the JWST pipeline with respect to near-infrared time-series observing modes.

## 2 Introduction

All detectors on board JWST make use of an up-the-ramp readout method, where multiple non-destructive reads of a given pixel are obtained as the charge within that pixel steadily accumulates. These ramps, more formally referred to as integrations, can then be repeated in a continuous sequence to construct a full exposure of a given astrophysical object or scene of interest. With multiple samples across each integration, signal "jumps" (commonly due to cosmic rays) can be identified for a desired user threshold, the onset of detector strong non-linearity or saturation can be identified through the ramp "levelling off", and partial ramps can still be used to estimate a signal rate. However, these advantages are not immediately realized, and care must be taken both when identifying jumps, and when measuring a signal rate from a given pixel ramp.

In the case of a jump, a quantifiable threshold or metric must be defined to assess whether a deviation of a single read within a ramp is truly discrepant, or simply within an expected range of variation. When fitting a ramp, decisions must be made to define the fitting algorithm itself, the relative weighting of individual points up the ramp, and the treatment of read noise. Both the jump detection and ramp fitting steps are further restricted by a need to perform these steps in a computationally tractable fashion while accounting for as many detector effects as possible. To ensure that the default processing for JWST data provides the most accurate flux rate measurements possible, a careful assessment of these decision points is necessary.

Throughout the first few years of JWST's operation, the default jump detection and ramp fitting algorithms have remained largely unchanged. The jump detection algorithm flags jumps using a two-point difference method (Anderson & Gordon, 2018) which compares the median STRtracted differences between reads (scaled by an estimate of the noise for each difference) against an input rejection threshold. However, additional modifications were made to better identify large cosmic-ray events like "snowballs" and "showers" (JWST-STScI-008545). For ramp fitting, a generalized least-squares method has been used (Fixsen et al., 2000) which weights the up-the-ramp samples on a pixel-by-pixel basis using an estimate of the pixel signal intensity. For pixels at high signal-to-noise regimes, reads at the beginning and the end of the ramp have higher weights than those in the middle, as the later samples have a lower fractional Poisson noise component. Conversely, the reads for pixels at low signal-to-noise regimes are weighted more uniformly, as the read noise on each read is more dominant. Nevertheless, a drawback of this ramp fitting process is that it relies on predefined signal-to-noise thresholds to define the weighting schemes, and cannot provide a more optimal continuous weighting scheme.

Recently, progress has been made to define a newer, more optimal, ramp-fitting routine that computes and leverages the covariance matrix between differenced reads (or resultants) to estimate the minimal $\chi^2$/maximum likelihood of the fit of a line to a given ramp (Brandt, 2024a). Typically, such approaches are computationally demanding in the frame of individual reads as a given read is correlated with all preceding reads, the resulting covariance matrix is dense, and inverting it to compute $\chi^2$ has a cost that scales as $n^3$, where $n$ is the number of reads. However, as non-overlapping differences between these reads do not share any photons, the covariance matrix is tridiagonal, and its inversion has a cost that scales as $n$. For more details on the algorithm to produce this covariance, we encourage the reader to review the original Brandt (2024a) publication. Once this covariance matrix is defined, jump detection can also be performed by comparing the $\chi^2$ value of a fit assuming all pixels are valid measurements to one that excludes differenced pixels at every possible jump location (Brandt, 2024b). These routines have already been applied to JWST imaging data in the aforementioned Brandt (2024a, 2024b) publications, and are now implemented as a non-default option of the official JWST pipeline.

With this newly developed method in hand, it is important to test and verify its suitability and capability across a wide range of JWST datasets and science cases. In this work we focus on the Near-InfraRed Imager and Slitless Spectrograph (NIRISS) / Single Object Slitless Spectrograph (SOSS), which is primarily used by the astronomical community for spectrophotometric time-series observations (TSOs) of transiting exoplanets. Such an investigation has particular importance for TSO observations, as the scientific community are already pushing the limits of JWSTs capabilities in an effort to detect ~part-per-million (ppm) wavelength dependent signals in the light curves of the lowest-mass, and potentially most Earth-like, exoplanets. At such precisions, the impacts of early reduction steps such as jump detection and ramp fitting become much more significant.

In Section 3 we outline our data collection and processing steps for both an initial on-sky dataset of WASP-39 b, and a suite of custom-made simulations. In Section 4 we present the analysis of the on-sky data, and in Section 5 we present the analysis of the simulated data. Finally, we provide our conclusions in Section 6.

# 3 Data Collection and Processing

## 3.1 On-Sky Data

For our on-sky data investigation, we focus on the Early Release Science (ERS) observation of the hot-Jupiter exoplanet WASP-39 b, with NIRISS/SOSS ([GO-1366](), PI: Batalha). This observation spans a total duration of ~8.2 hours, across 537 integrations, and captures the transit of WASP-39 b as it passes in front of its star relative to our line of sight. Using the GR700XD grism and the SUBSTRIP256 subarray, both Order 1 and Order 2 spectra are measured, spanning a combined wavelength range of ~0.6-2.8 μm. Order 3 also falls within the subarray, although due to its comparably low counts (peak flux of ~3% of Order 2) and current lack of formal support, we exclude it from our investigation. However, we do note that Order 3 will be calibrated in the future as part of [GO-3279](), PI: Hoeijmakers.

These data have already been extensively analyzed as part of the broader ERS effort, resulting in an initial publication by Feinstein et al. (2023). This work revealed the presence of $H_2O$ and K absorption features, as well as signatures of clouds, but also provided early refinements to NIRISS/SOSS-specific reduction and analysis processes. Recently, Carter & May et al. (2024) additionally presented a joint reduction of these NIRISS/SOSS data in conjunction with similar observations using NIRSpec PRISM, NIRSpec G395H, and NIRCam F322W2. While the reduction processes remained largely similar between this work and that of Feinstein et al. (2023), an improved joint-fit of the extracted white light curves across these four observations, in addition to archival TESS and NGTS data, provided sub-percent constraints on WASP-39 b's orbital parameters.

In totality, this dataset is an ideal choice for this investigation. The data are high quality, with no evidence for significant uncorrected systematics that may complicate our analysis, and the orbital parameters have already been tightly constrained, expediting our ability to explore the impacts of the jump detection and ramp fitting algorithms on "final-level" transmission spectra.

## 3.2 Simulated Data

For our simulated data investigation, we utilize SOSS simulations as generated by Loïc Albert

(Université de Montréal) using IDT-SOSS (Albert et al. 2023, briefly presented in section 9). IDT-SOSS uses WebbPSF to model the trace profile at a large number of wavelengths to seed SOSS spectral traces (orders 1, 2, 3 as well as 0 and -1) on 2D detector images. It uses the pre-Launch wavelength calibration, spectral trace positions and photon throughput to set the exact trace positions and intensity. A grid of model stellar atmospheres at R > 250000 is used as an input covering 3.0 < log(g) < 5.0 and 2300 K < Teff < 6900 K (P. Hauschildt, priv. comm.). Also, a limited set of planetary atmosphere models are provided by Björn Benneke (priv. comm.). IDT-SOSS uses the formalism of Mandel & Agol (2002) to model the transit light curve. A detector image is created at each integration time step. IDT-SOSS can add most of the known sources of noise to images (readout, photon noise, flat field errors, non-linearity, etc) but here, we decided to produce a noiseless simulation to limit the sources of possible errors in the investigation. The stellar atmosphere models tabulate specific flux at >60 radial points across the disk that can be interpolated and fit to model wave-dependent limb darkening following the predictions of Claret et al. (2000). But for this simulation, we opted to remove limb darkening (produce a trapezoidal transit light curve), again to prevent the introduction of errors due to the chosen limb darkening laws.

For this experiment, we used a star with Teff = 5400 K, $\log(g)$ = 4.5 to mimic WASP-39, and a synthetic planet with a uniform transit depth of 1% at all wavelengths. The simulation was generated at twice the pixel resolution to minimize digitalization noise and then binned down to the native pixel resolution. The stellar model was scaled to a 2MASS *J*=10.663, to match WASP-39 b, and we used $N_{groups}$=9, and $N_{integrations}$=537 for a duration of 8.18 h centered on the transit. The orbital period adopted was 4.0552842 days, the impact parameter was 0.4498, $\sqrt{e} \sin \omega$ = 0.071, $\sqrt{e} \cos \omega$ = 0.071, all adopted from Carter & May et al. 2024. To facilitate integration with the JWST pipeline, these simulated ramps are injected directly into the `*uncal.fits` files of the real WASP-39 b SOSS observation described in Section 3.1.

We define and apply five separate noise cases to the simulated data: "Pristine", "Poisson", "White", "1/f", and "All". For the Pristine case, we do not add any additional noise and keep the 2D spectral time-series in its original state. For the Poisson case, we directly add Poisson noise to each pixel of each group using the `random.poisson()` function in `Numpy`. For the White case, we add white noise to each pixel of each group using the `random.normal()` function in `Numpy`, with the standard deviation $\sigma_w$ = 6.58. For the 1/f case, we construct a time-series of $1/f$ noise using the `processes.noise.ColoredNoise()` function included in `stochastic`, with the exponent $\beta$ = 1.0217. When constructing this time-series we adopt a 10 $\mu s$ time between pixel reads (i.e. the rate at which $1/f$ is applied to the data), and a 120 $\mu s$ gap between the last read of a given column and the first read of the next column. The variance of the $1/f$ time-series is also rescaled by a factor $\sigma_f$ = 6.11 to match the amplitude seen in on-sky data. Finally, for the All case we combine the existing Poisson, White, and 1/f cases together.

The values for $\sigma_w$, $\beta$, and $\sigma_f$ are determined by applying Approximate Bayesian Computation (ABC) to a sample of SOSS dark frames obtained during commissioning for SUBSTRIPE96. While the data explored here use the SUBSTRIP256 subarray, we do not expect the results between subarrays to vary significantly as the dark signal is independent of the frame time, and it was more

JWST-STScI-008975
Revision -
computationally efficient to run the analysis on a smaller subarray. Specifically, we begin with the group-level darks and use the implicit time-stamps for every pixel within that frame to compute a 1D time-series for each individual group. These are cleaned for outliers, and a power spectral density (PSD) is then constructed for each group, which we median stack, see Figure 1 (Left). While significant substructure is present, it is clear that the power spectrum follows a linear slope at lower frequencies, indicative of red/pink $1/f$ noise, before hitting a floor at the higher frequencies, indicative of white noise. To infer the parameters of this process, we use the abeec package (https://github.com/nespinoza/abeec) following a methodology similar to that outlined in Witzel et al. (2018), where we generate a realization of the same number of frames as our input data using a sample from several realizations of $1/f$ processes with different $\sigma_w$, $\beta$, and $\sigma_f$ values using the stochastic Python library. Then, we compute the median PSD in the exact same way as our data, and compute the distance between our simulated data and the real data as the absolute difference of the logarithm of the simulated and real PSD's (scripts available at https://github.com/nespinoza/abeec/tree/master/examples/darks-example). The results of our posterior samples for $\sigma_w$, $\beta$, and $\sigma_f$ are shown in Figure 1 (right).

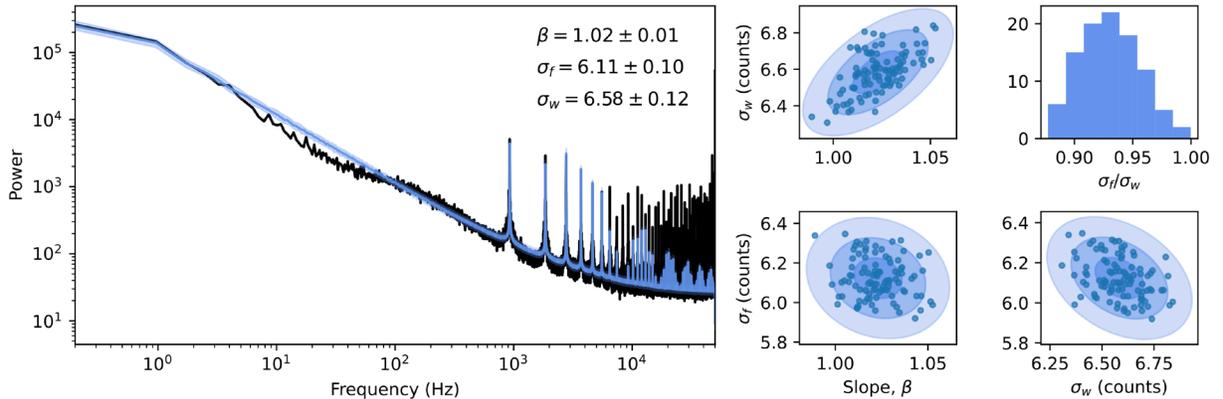

*Figure 1*: **Best fit power spectral density (PSD) determined from approximate Bayesian computation.** *Left*: **The best fit power spectral density (blue), alongside the measured PSD from NIRISS/SOSS dark frames.** *Right*: **Posterior distributions across the three fit parameters of the white noise component ($\sigma_w$), 1/f noise component ($\sigma_f$), and the slope ($\beta$).**

## 3.3   Stage 1 Detector Processing

We begin our reduction with the lowest level, `*uncal.fits`, files publicly available in the Mikulski Archive for Space Telescopes (MAST). For both on-sky and simulated data we make use of a custom Python implementation of the JWST pipeline which uses `CRDS_VER=11.17.19`. Due to differences in the time the pipeline was applied, the on-sky data uses `CRDS_CTX=jwst_12.41.pmap`, whereas the simulated data uses `CRDS_CTX=jwst_13.03.pmap`. As the on-sky data served primarily as an initial investigation to motivate the suite of simulated data, and that similar results can be observed across them, it is unlikely that this has significantly affected our conclusions.

For the on-sky data, all data undergo: data quality initialization, saturation detection, superbias subtraction, reference pixel correction, linearity correction, and dark current subtraction. In contrast, for the simulated data we only perform the data quality initialization, as all of the other

Check with the JWST SOCCER Database at: https://soccer.stsci.edu
To verify that this is the current version.

listed steps are not included in the simulation framework and will only serve to introduce additional noise into the data. Conceptually, this is equivalent to the on-sky case of these steps being performed perfectly, and the effects of jump detection and ramp fitting being isolated. Irrespective of dataset, our reduction approach for jump detection and ramp fitting diverges into two.

For our first reduction approach, hereafter referred to as "default pipeline", we continue the reduction through the default JWST pipeline jump detection and ramp fitting steps, in essence producing identical Stage 1 `*rateints.fits` products to those available on MAST. For our second reduction approach, hereafter referred to as "likelihood", we substitute both the default jump detection and ramp fitting steps of the JWST pipeline with the optional likelihood-based methods described in Brandt (2024a, 2024b) by setting `alogrithm='LIKELY'` in the input parameters of the ramp fitting step. We note that, at present, this does not need to be applied to the jump step, as by enabling its use during ramp fitting the routine will disregard any existing jump flagging and perform the likelihood-based flagging within the ramp fitting step itself.

### 3.4 Stage 2 Spectroscopic Processing

Following the production of `*rateints.fits` equivalent products for the two reduction approaches, we apply an identical set of routines to further calibrate the data and extract the spectroscopic time-series. While some of these routines utilize the JWST pipeline, many make use of custom routines developed by the NIRISS/SOSS team at STScI.

Before any calibration occurs, we first apply the default JWST pipeline to assign a world coordinate system (WCS) object with each exposure. For SOSS the WCS is not useful for the on-sky positions, as it is a dispersed spectrum, but this step is still valuable as it applies a wavelength map to the data. Then, we perform a correction for bad pixels on each 2D image using the `interpolate_replace_nans()` function included in `astropy`, with a 2D Gaussian kernel of $\sigma_x = 3, \sigma_y = 3$.

For on-sky data, NIRISS/SOSS observations exhibit a spatially varying background flux that must be subtracted to enable an accurate spectral extraction. The background is difficult to estimate within a science image, and as such we subtract it using a scaled empirically defined background template, measured during commissioning. To estimate the appropriate scale factor, we extract the median of a box of pixels that spans the sharp background transition ($x = 600$ to $800$, $y = 210$ to $250$; see Figure 2) and take the ratio compared to a similar median from the empirical template. This is performed on an integration-by-integration basis, and a final median of these ratios is calculated to estimate a single scale factor for the entire observation. The background signal is not included in the simulated data, and therefore is not corrected.

JWST's near-infrared detectors all exhibit $1/f$ noise, correlated noise resulting from biases in the detector readout electronics. For NIRISS/SOSS, this is visible in the form of positive or negative "striping" between columns across the subarray, which vary between individual groups of the up-the-ramp readout. However, due to the compounding effects of $1/f$ noise and the NIRISS background, we perform the subtraction of the $1/f$ noise at the integration level and leave further improvements to this correction for future work. To estimate the $1/f$ noise to subtract from each integration, we first calculate a median image across all integrations, and then set any pixels greater than 0.8% of the peak of the median background-subtracted image to NaN. This threshold acts to suppress the flux from the spectroscopic traces, and leave behind the regions of the image that are primarily influenced by $1/f$ noise. This median image is then subtracted from each integration. Finally, the $1/f$ noise is estimated and removed using a column-wise median of this subtracted image. We note that the $1/f$ noise is only present for the on-sky data, and the 1/f and All noise prescriptions of the simulated data – this correction is not applied for the other noise prescriptions.

Prior to spectral extraction, a flat-field division and pathloss correction are applied using the default JWST pipeline. The spectral traces are obtained directly from the publicly available `PASTASOSS` Python package (Baines et al., 2023), using the corresponding pupil-wheel position for this observation. Finally, the wavelength dependent spectroscopic time-series is extracted using a custom box extraction centered on these traces, with an aperture half-height of 15 pixels. We note that while specific decisions on parameters like the aperture half-height (or others previously mentioned) may influence the measurements from a given observation, the primary interest of this report is in *relative* differences between different reductions of the same dataset. As a result, we do not perform any significant investigation or optimization of these parameters.

### 3.5 Light Curve Fitting

We use the publicly available, Python-based, exoplanet time-series reduction and analysis pipeline, `Eureka!`(Bell et al., 2022) to extract and fit the spectrophotometric light curves from this data. The light curves are generated by binning the spectroscopic time-series along the wavelength axis, using the optimal spectral bins defined in Carter & May et al. (2024), for a resolving power of $R = 100$.

As previously mentioned, the on-sky data have been heavily investigated alongside NIRCam, NIRSpec, TESS and NGTS light curves. Therefore, we do not perform any fitting to the white light curves from Order 1 and 2, and instead adopt the best-fit orbital parameters as reported in Carter & May et al. (2024). Specifically, for the on-sky fitting we fix the period, $P = 4.055284$ d, the transit mid-time, $t_0 = 59787.0567843$ MBJD TDB, the inclination, $i = 87.7369°$, and the scaled semi-major axis, $a = 11.390$. We also assume an eccentricity, $e = 0$, and an argument of periapsis, $w = 90°$. The limb darkening followed a quadratic law, with the parameters estimated using `ExoTiC-LD` across a 3D stellar model grid, for an effective temperature, $T_{eff} = 5512$ K, a metallicity, $[M/H] = 0.0$, and a surface gravity, $\log(g) = 4.7$.

When performing fits to the simulated data using matching orbital parameters to the on-sky data, we observed wavelength-dependent residuals in the light curve fits that restricted our ability to interpret the transmission spectra, primarily due to differences in the transit central time and limb darkening. As we are primarily interested in differences between different reductions of the same datasets, and to avoid the time cost repeating our simulation, we perform a white light curve fit on the Pristine data to refine the best fit parameters. In this process we keep the transit depth fixed at the known value of 1% and the limb darkening parameters fixed to 0, and find best fit values for $P = 4.0552842$, $t_0 = 59787.03637913$, $i = 87.73611$, $a = 11.43161$, $e = 0.00422$, and $w = 39.57747°$. Upon performing spectrophotometric light curve fits to the Pristine data with these fixed parameters we find an excellent agreement with the simulation input, with all transit depths closely distributed around 1% with a scatter below 1 ppm (see Figure 8 in Section 5). Therefore, despite the differences to our on-sky fits, we are confident they do not impact the relative measurements between the two methods and fix the parameters to these values for all further simulated data light curve fits.

When performing the spectrophotometric light curve fits to both the on-sky and simulated data we use a first-order polynomial in time as the only systematics model, and the astrophysical transit model is computed using `batman` (Kreidberg, 2015). The actual fitting of the transit depth and systematics model parameters are performed using Markov Chain Monte Carlo (MCMC) through the publicly available Python package `emcee` (Foreman-Mackey et al., 2013), using 200 walkers, across 1100 steps, with the first 100 steps discarded as burn-in.

## 4   Analysis – On-Sky Data

Beginning with the on-sky data, we assessed differences at the Stage 1, `*rateints.fits`, level as shown in Figure 2. At this stage, no $1/f$ or background correction has been applied to the data and we can visualize differences immediately following jump detection and ramp fitting. The curved Order 1, 2, and 3 traces (bottom, center right, upper left) are immediately obvious, alongside the spatially structured background (sharp transition at pixel column ~700), and various Order 0 contaminating sources. White pixels correspond to pixels masked by the data quality initialization step. Differences are not visible by eye between the methods, but become immediately obvious when taking the residuals between them. A distinct vertical banding structure is present throughout the image, and is indicative of a difference in impact of $1/f$ noise between the two jump detection and ramp fitting methods. Pixels in the background regions away from the traces are impacted fairly similarly, but significant and correlated deviations are present across the Order 3 trace, at the positions of Order 0 contaminants, and across the wings of the Order 1 and 2 traces. Conversely, the differences along the Order 1 and 2 traces appear less correlated. In combination, it is clear that the difference between the two methods for a given pixel is significantly impacted by its underlying flux level and $1/f$ noise in a non-linear fashion.

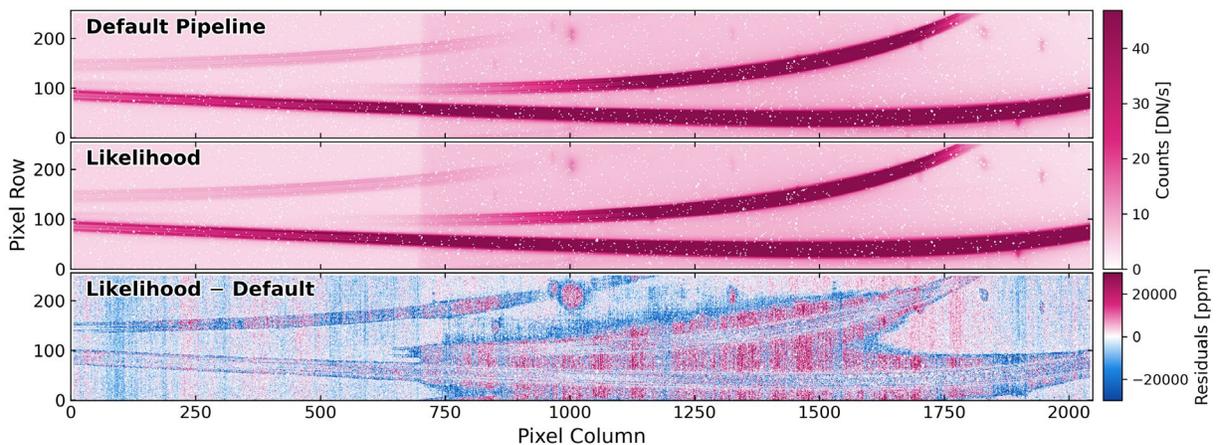

*Figure 2*: rateints.fits level images for the NIRISS/SOSS observation of WASP-39 b. *Top*: Image using the default pipeline jump detection and ramp fitting. *Middle*: Image using the likelihood based jump detection and ramp fitting.

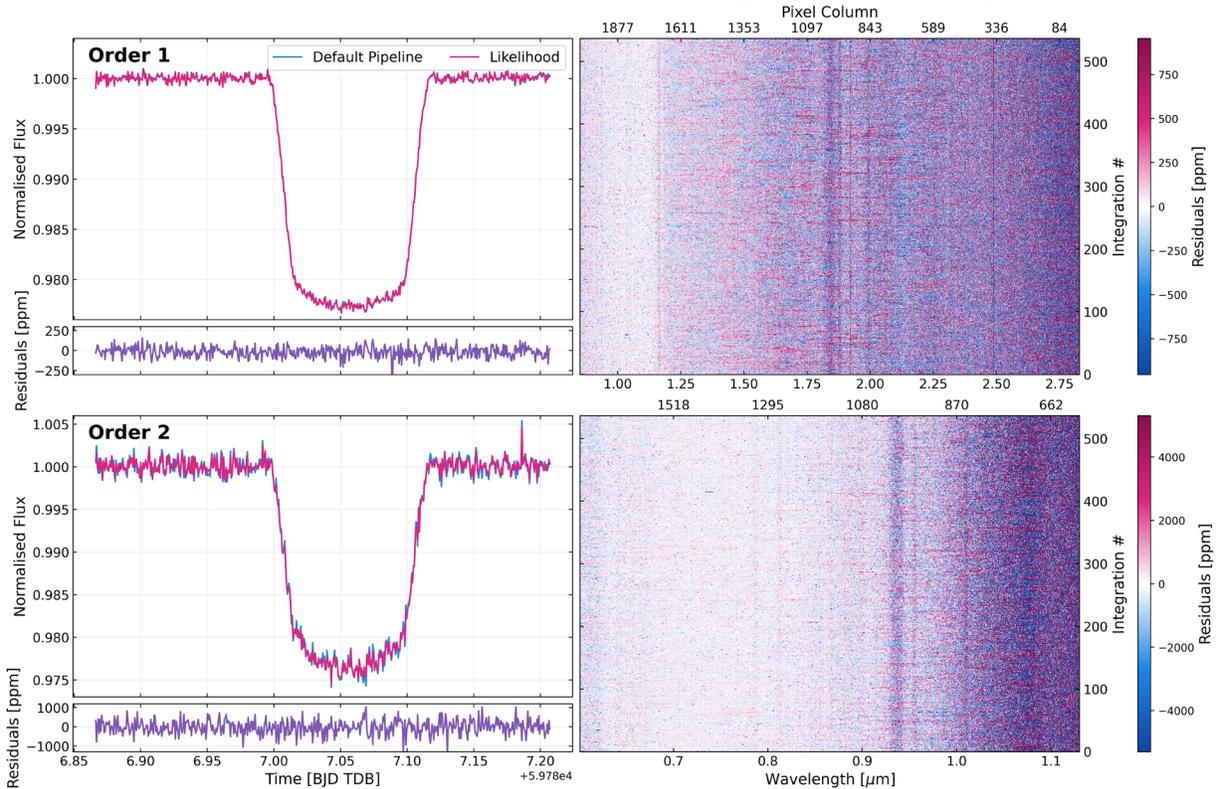

*Figure 3*: Light curves for the NIRISS/SOSS observations of WASP-39 b. *Left*: Order 1 and 2 white light curves for both the default pipeline (blue) and likelihood (pink) jump detection and ramp fitting methods. *Right*: Residuals of the 2D spectroscopic light curves between the two methods.

Following further data calibration and light curve extraction, differences are still present between the two methods as shown in Figure 3. While the two methods broadly produce the same structure in the white light curves, there is less apparent noise for the likelihood method. This is more obvious for the Order 2 light curve, which has a residual standard deviation of ~370 ppm, versus ~68 ppm for the Order 1 light curve. We also show the difference between the two methods in the form of 2D wavelength-dependent light curves. For both Order 1 and Order 2, differences are strongest at towards the lowest illumination areas of the traces, indicative of Poisson noise as the dominant modifying factor. Distinct vertical banding across some pixel columns is also present, and lines up with the presence of Order 0 contaminants visible in Figure 2. In this case it is likely that the presence of these sources has biased the adopted $1/f$ cleaning routine Finally, sparse horizontal striping that varies from integration to integration is present, and is driven by the differential interactions of $1/f$ noise with the ramp fitting methods as seen in Figure 2 (bottom).

The differences shown thus far provide valuable insights, but most important are the differences in the final extracted transmission spectrum shown in Figure 4. Here we see that while differences are present, they can be subtle, with a standard deviation in the residuals between methods of ~50 ppm. At the most illuminated wavelengths, differences are less apparent, but these differences grow more significant towards the longer wavelengths. Whether these difference are significant enough to bias the current interpretation of WASP-39 b's atmosphere is beyond the scope of this report, however, it is useful to consider their broader implications. WASP-39 b has a high (~1200 K) temperature, large (~2%) transit depth, and orbits a relatively bright ($J$~10.6 mag) star, which combine to make it an ideal and easy target to observe with JWST. For a more challenging target with more subtle atmospheric signatures, these residuals are likely to be more significant, especially as many observers are already targeting ~10's of ppm level signals (e.g. Piaulet-Ghorayeb et al., 2024; Cadieux et al, 2024; Lim et al., 2023; Madhusudhan et al., 2023).

In total this on-sky analysis demonstrates that the differences between the default pipeline and likelihood methods are present at every stage of the reduction, and exist at a level that may be significant for many current and future observations. However, as an absolute understanding of WASP-39 b's atmosphere is not known, this analysis cannot determine which method provides the most accurate answer, or whether the accuracy is globally improved when using one method over another.

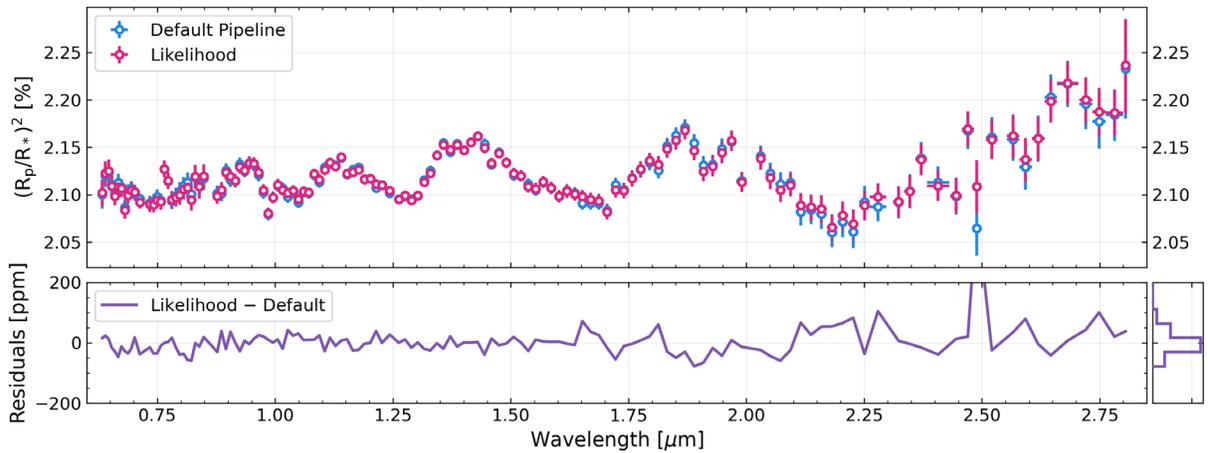

*Figure 4:* **Extracted NIRISS/SOSS transmission spectrum for WASP-39 b using both the default pipeline (blue) and likelihood (pink) jump detection and ramp fitting methods.**

## 5 Analysis – Simulated Data

### 5.1 2D Spectra

To understand not just the relative differences between the jump detection and ramp fitting methodologies, but also their absolute differences versus a known ground truth, we must turn to simulated data. The `*rateints.fits` level simulated images for each of the five separate noise cases described in Section 3.2 ("Pristine", "Poisson", "White", "1/f", and "All") are shown in Figure 5.

As evident from the on-sky data, it is difficult to visualize the noise components in the 2D spectra directly. However, for each of simulated noise cases we can subtract off the Pristine images to isolate differences at the `*rateints.fits` level that result from the influence of the noise on the jump detection and ramp fitting. In Figure 5 we see the expected structure for the Poisson case, where brighter pixels exhibit a larger absolute noise level, and measure no median offset relative to the Pristine image (<< 1ppm). For the White case we would expect broad uncorrelated noise across the full image, but instead we see that the noise is slightly elevated along the spectral traces where the flux is higher. As white noise is explicitly applied in an additive manner, and *not* a multiplicative one, this correlated structure must be a result of the jump detection and ramp fitting. For the 1/f case we see the expected column-wise noise structure, but also observe correlated noise along the traces that in many columns is of a higher amplitude than the noise outside of the spectral trace for that column. As with the White case, the $1/f$ noise is applied in an additive manner, so this structure must result from the jump detection and ramp fitting. For both the White and 1/f cases, we measure a positive median offset relative to the Pristine image of ~400 ppm. Finally, for the All noise case we see that each of the structures are combined, and the correlated noise along the traces from the White and 1/f cases is much less apparent alongside the Poisson noise. Interestingly, the median offset for the All noise case is lower than the White and 1/f cases, at ~250 ppm.

In the right-hand side of Figure 5 we isolate the differences between the jump detection and ramp fitting methods. Across all integrations for the Pristine case we see that the differences are close to zero ($\sigma \approx 2$ ppm; mean offset <1ppm), which shows that in the absence of noise the two methods perform largely identically. For the Poisson case, we see there are differences along the traces, but their amplitude is correlated with the underlying flux level. The mean offset across all integrations is relatively small at ~2 ppm, and the standard deviation is ~12,000 ppm, indicating a very small preference towards lower fluxes for the default method. The White noise behaves similarly to the Poisson case, with differences correlated to the flux levels and a standard deviation of ~10,000 ppm, however, the mean offset is much larger in amplitude at ~−410 ppm. For the 1/f case the differences also seem to be correlated with the underlying flux levels, but are also strongly correlated along the detector columns. While by eye the amplitude of the differences might seem more distinct, the standard deviation is ~10,000 ppm, and is therefore comparable to the other noise sources. The mean offset is less descriptive given the correlated nature of the $1/f$ noise, but it is slightly lower than the White noise case at ~−350 ppm. Finally, for the All case we see the noise sources combined, and measure a mean offset of ~−790 ppm with a standard deviation of ~29,000 ppm. We also note that for all cases the median offset across all integrations is equal to zero, indicating an equal distribution of positive and negative offsets.

The impact of these offsets is best considered in terms of their impact of an extracted spectrum. The cross-dispersion of the SOSS orders is closely aligned to the columns (Filippazzo report in preparation), meaning that differences are not as significant if they are *evenly* distributed around zero for a given column. For the White case this is not true, as the mean offset is negative, and will act as a dilution for the default pipeline method versus the likelihood method. However, for the 1/f and All cases the effects are far more significant due to the column-by-column offsets, and will lead to wavelength-correlated biases between the two methods.

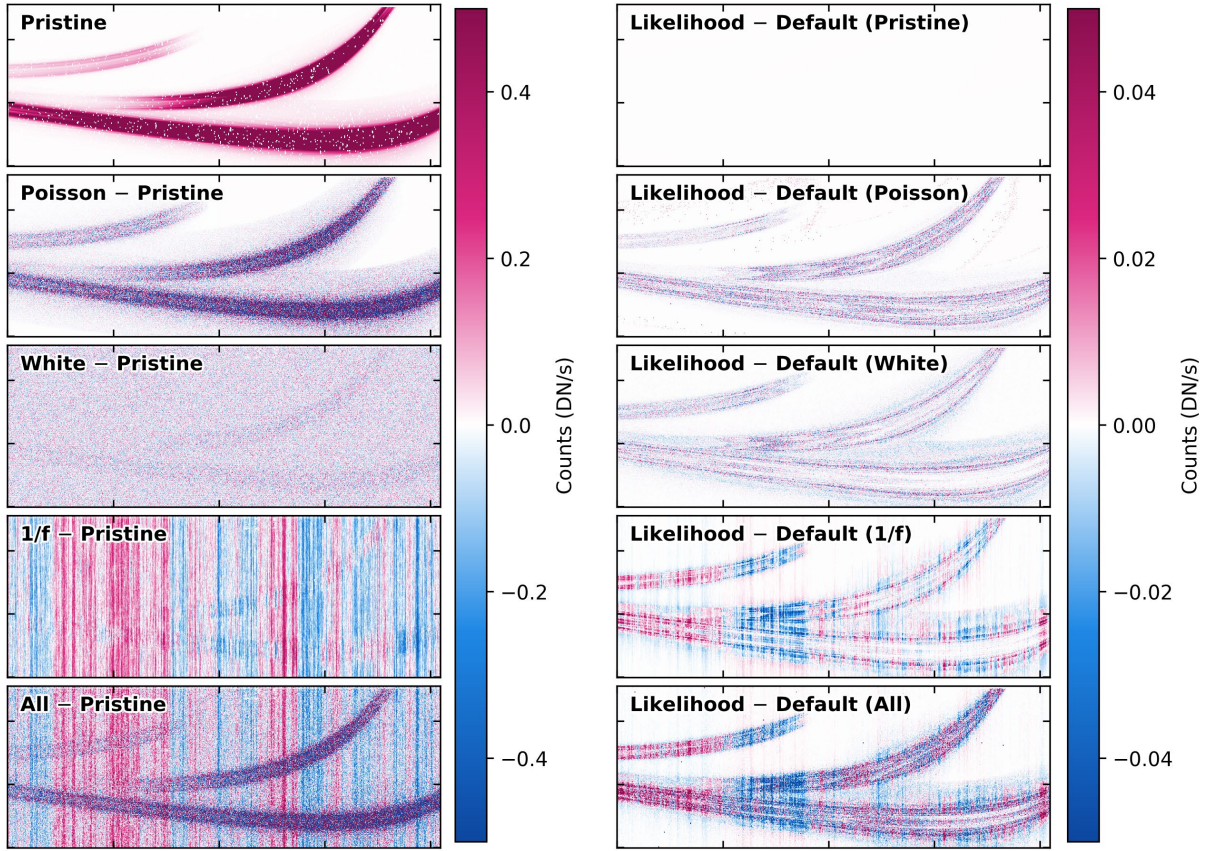

*Figure 5*: **Simulated *rateints.fits 2D spectra and comparisons, for the Pristine, Poisson, White, 1/f, and All noise cases described in Section 3.2.** *Left*: **A single integration for the Pristine noise case (top), alongside matching Pristine subtracted integrations for other four noise cases.** *Right*: **Difference images for all noise cases between the default pipeline and likelihood jump detection and ramp fitting methods.**

## 5.2  $1/f$ Noise Correction

While $1/f$ noise can drive significant variations both across and between methods, there are strategies to mitigate its impact prior to spectral extraction. Broadly speaking, these aim to measure and subtract the noise in a column-wise manner, perhaps through a median or linear fit, using pixels in "unilluminated" regions of the subarray. Ideally this correction should be applied at the group-level, as $1/f$ noise originates during detector readout, however, this may be skipped in lieu of an integration-level correction for efficiency, or to reduce complexity. For example, $1/f$ noise in on-sky SOSS data compounds with the spatially correlated background (Baines et al., report in preparation), and necessitates a careful handling of both components at the group-level which has been broadly adopted by the wider community (e.g. Radica et al., 2025; Piaulet-Ghorayeb et al., 2024; Lim et al., 2023). The adoption of such an approach as a default procedure for the official JWST pipeline is still under development and consideration, and as such we rely on the simpler integration-level correction described in Section 3.4.

The Pristine-subtracted residuals for the 1/f and All noise cases are displayed in Figure 6. For the 1/f case, two distinct residual components are present: 1) higher frequency residuals across columns, indicative of the median subtraction not capturing variations of the $1/f$ noise along a column, and 2) strong residuals along the spectral traces, driven by the differences in flux estimation imprinted by the ramp fitting process (also seen in Figure 5). Improving 1) requires improving the $1/f$ correction method itself, however, 2) is an explicit result of the ramp fitting step – the presence of $1/f$ noise biases the estimated flux more strongly in more highly illuminated pixels – and can only be improved with a group-level subtraction. Turning to the All noise case, the presence of the Poisson and white noise makes both 1) and 2) difficult to distinguish, and by eye the correction looks reasonable. In reality, the column-wise correlated residuals are still present, and will bias the spectral extraction.

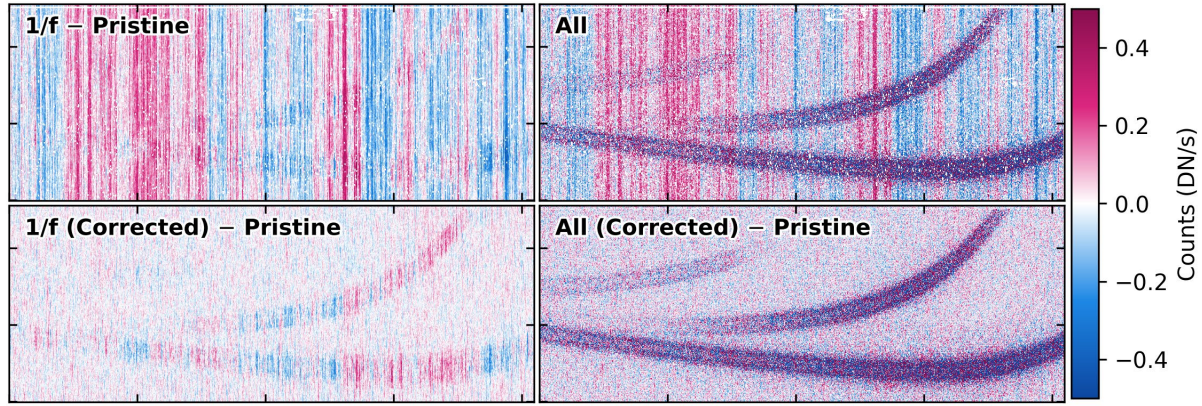

*Figure 6*: Pre- and post-$1/f$ correction *rateints.fits 2D spectra for cases with only $1/f$ noise, and all considered noise sources.

## 5.3  Transmission Spectra

Following the $1/f$ correction, spectral extraction, and light curve fitting, we produce transmission spectra for all of the simulated noise cases as shown in Figure 7 alongside a variety of statistical quantities in Table 1.

For the Pristine case the results are incredibly consistent with the grey 1% injected transit for both the default and likelihood methods, and their differences are in strong agreement, with a ~1 ppm scatter. Upon addition of Poisson noise, a higher level of wavelength correlated structure is introduced, but the median offsets are raised only slightly to ~5 ppm and the differences display a ~2 ppm scatter.

For White noise, there is an observable preference for both methods to fit slightly higher transit depths towards longer wavelengths (lower signal) although the median offsets and differences are approximately equivalent to the Poisson case at ~6 ppm and ~2 ppm scatter. Given that the distribution of the residuals appears symmetric, this indicates a shared effect that biases results to larger transit depths in the presence of white noise.

In the 1/f case, similar amplitude wavelength correlated residuals are introduced for both methods, although the differences between them become more distinct. Across the vast majority of wavelengths, the likelihood method is closer to the ground truth 1% transit, and the amplitude of the differences are anti-correlated with the flux at that wavelength. Importantly, we see a ~11 ppm scatter between the two methods, with maximum deviations of ~79 ppm.

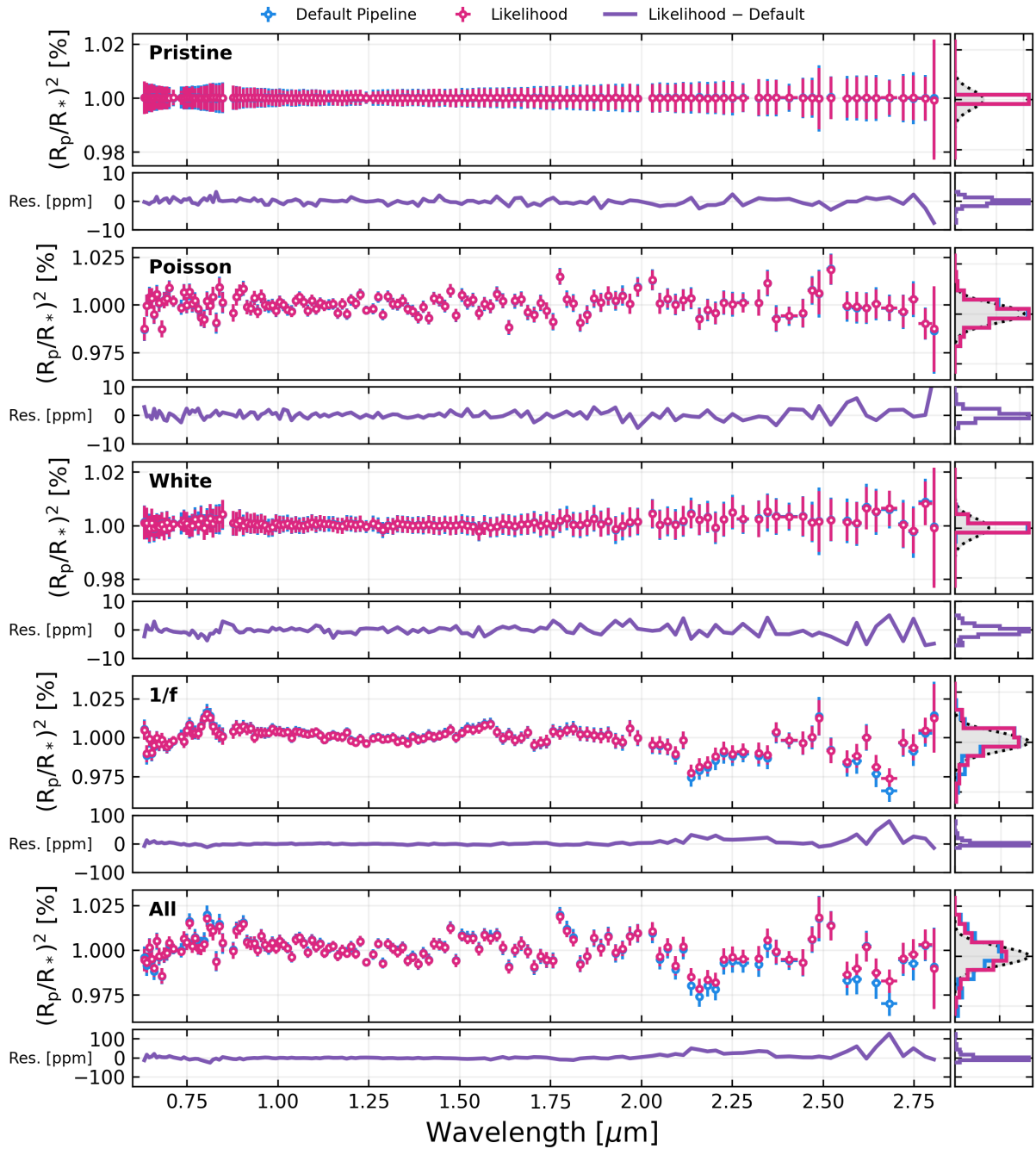

*Figure 7*: Transmission spectra for the Pristine, Poisson, White, 1/f, and All noise cases (top to bottom), for both the default (blue) and likelihood (pink) methods, alongside their transit depth distributions relative to the expected Gaussian noise distribution (top right panels). Also shown are the residuals between the two methods (purple, bottom left panels) and their distributions (bottom right panels).

Finally, for the All noise case we see the effects of all of the noise components combined together, displaying a wavelength-correlated structure that appears to match a combination of the scatter from the Poisson and $1/f$ noise components. Here the median offsets are ~12 ppm for both methods, although the scatter between them is elevated to ~17 ppm, with maximum deviations of ~130 ppm. This is ~3 times less than the scatter observed for the on-sky data, and may indicate that additional factors excluded from the simulations, such as detector non-linearity, charge migration, and/or cosmic rays, may have a significant differential impact on ramp fitting method. Further investigation into these effects is worthwhile, but the underlying result for these simulations is clear – the likelihood method produces more accurate results. While there are some cases where it produces measurements further away from the 1% ground truth than the default method, we consistently find that it will instead provide a better agreement to underlying Poisson case, which is a more accurate representation of the true measurements from the simulated observations.

The remaining wavelength-correlated structure in the All noise case has the potential to be misleading when interpreting atmospheric signals. A large fraction of this structure comes from the $1/f$ noise, however, a significant level of variation is still present from the other noise sources. Assuming we can combine the noise sources linearly, even with an optimal $1/f$ noise correction there would still be a ~6 ppm scatter in the differences between the two methods, with maximum deviations of ~50 ppm. Such differences may be significant enough to bias the interpretation of SOSS transmission spectra, and in isolation the likelihood method should be considered a more optimal choice than the current pipeline implementation. The Poisson case also serves as an important reminder that even for an ideal systematics correction, transmission spectra can easily exhibit "by-eye" wavelength dependent structures that should not replace statistically motivated model fits and conclusions.

|  | Pristine | Poisson | White | 1/f | All |
|---|---|---|---|---|---|
| ***Single Simulation*** |  |  |  |  |  |
| Median Offset (Default) [ppm] | 0.83 | 5.3 | 6.3 | 7.3 | 12 |
| Median Offset (Likelihood) [ppm] | 0.72 | 5.7 | 5.8 | 5.5 | 13 |
| Median Difference [ppm] | -0.046 | 0.10 | -0.32 | -0.84 | -0.79 |
| STD Difference [ppm] | 1.2 | 1.9 | 1.8 | 11 | 17 |
| Max Difference [ppm] | 7.6 | 14 | 5.5 | 79 | 130 |
| ***Multi Simulation*** |  |  |  |  |  |
| Median Offset (Default) [ppm] | $0.82^{+0.76}_{-0.81}$ | $1.3^{+46}_{-46}$ | $5.9^{+16}_{-8.3}$ | $2.3^{+52}_{-47}$ | $7.5^{+71}_{-64}$ |
| Median Offset (Likelihood) [ppm] | $0.78^{+0.73}_{-0.70}$ | $1.2^{+46}_{-46}$ | $5.6^{+17}_{-8.1}$ | $2.2^{+48}_{-43}$ | $7.8^{+66}_{-60}$ |
| Max Offset (Default)[ppm] | 6.4 | 530 | 160 | 840 | 990 |
| Max Offset (Likelihood) [ppm] | 9.5 | 540 | 150 | 740 | 820 |
| STD (Default) [ppm] | 0.91 | 52 | 17 | 82 | 99 |
| STD (Likelihood) [ppm] | 0.89 | 52 | 17 | 72 | 86 |

*Table 1*: Statistical quantities measured from the transmission spectra for all five noise cases, for both single and multi-simulation scenarios.

## 5.4 Multi-Simulation Noise Properties

Although the variety of simulated noise cases allows us to determine the relative performance of the two jump detection and ramp fitting methods, a single noise realization may be misleading when assessing the overall distribution over transit depth. Therefore, we repeated the entire simulation framework, stopping after 23 independent realizations were made. The specific number of realizations was not carefully selected, and was primarily limited by available computing resources. We then calculated the overall transit depth distributions across all realizations, analogous to the top-right panels of Figure 7, and show them in Figure 8.

The distributions closely match those for the individual case presented in Section 5.3, and provide some additional information on the interactions of different noise sources on transmission spectra. The Pristine example is tightly constrained to 1%, as expected for no noise, and is similarly inconsistent with the assumed error bars. The Poisson case provides too tight of a distribution compared to the expectation from the estimated transit depth errors, suggesting that the uncertainties are slightly overestimated by the light curve fitting process.

The White noise case matches the earlier findings that it preferentially biases the fit to deeper transit depths, particularly for less illuminated regions of the detector. In this context, for either available jump detection and ramp fitting method, white noise acts as a wavelength dependent inverse dilution effect and can potentially mimic broad absorption features in transmission spectra. From the performed simulations, we find a maximum deviation of ~150 ppm for both methods.

For the 1/f case, we see that the noise distribution changes quite distinctly, and does not match the underlying assumption for Gaussian noise, with broader wings and a slight asymmetry. In this case we see much larger offsets of ~840 and 740 ppm for the default and likelihood methods respectively, alongside a difference in the standard deviation of ~82 and ~72 ppm. This shows that both in terms of typical and maximum deviations from the ground truth, the likelihood method offers a ~12% improvement over the default method.

Finally, for the All noise case, the noise distribution matches a combination of the Poisson, White, and 1/f distributions, and is distinctly non-Gaussian and asymmetric. While we do not investigate in detail the impact of covariances between noise sources, this result indicates that their effects are not as significant as their isolated effects. The maximum offsets remain close to ~10% at 990 and 820 ppm for the default and likelihood methods respectively, along with standard deviations of ~1% at 99 and 86 ppm. Similarly, to the 1/f case, the likelihood method offers a ~18% improvement in maximum deviations from the ground truth, and a ~13% improvement in the typical deviations.

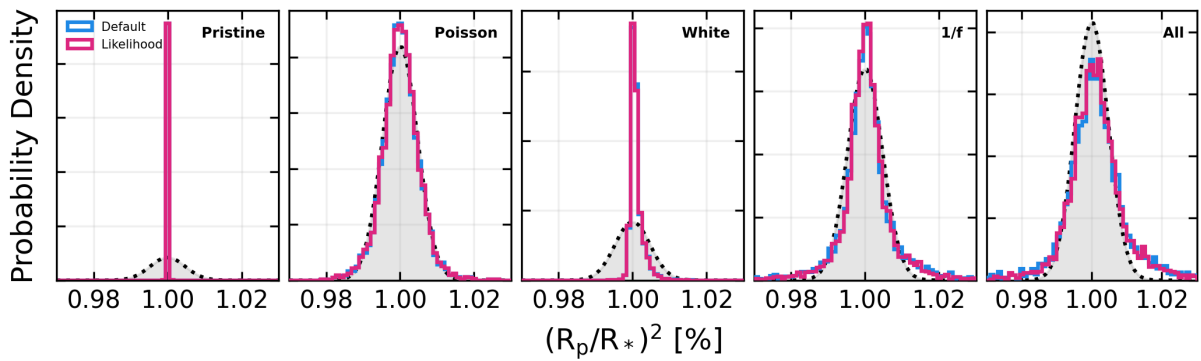

*Figure 8*: Combined transit depth distributions across 23 independent simulations of the Pristine, Poisson, White, 1/f, and All noise cases (left to right), for both the default (blue) and likelihood (pink) jump detection and ramp fitting methods, alongside the expected distributions for Gaussian noise (black/grey).

# 6 Conclusions

In this work we explored the differential effects of jump detection and ramp fitting procedures, particularly the present default JWST pipeline and the recently implemented "LIKELY" method described by Brandt (2024a, 2024b), with a focus on exoplanet time-series observations with NIRISS/SOSS. After analysis of both on-sky data, and a suite of simulated data with different input noise prescriptions, we have reached the following conclusions:

- Differences between the default and likelihood jump detection and ramp fitting methods are observed for on-sky data of WASP-39 b, and result in a residual scatter of ~50 ppm between measured transmission spectra. The relative accuracy of the two methods for the on-sky data cannot be assessed as the true transmission spectrum cannot be known, however, this scatter is significant enough to impact a range of JWST time-series observations targeting atmospheric signals of ~10's of ppm or lower.

- Following analysis of a suite of simulated data, across a range of input noise scenarios, the likelihood method is demonstrated to provide a more accurate measurement of the injected ground truth transmission spectrum. For the cases with only $1/f$ noise, and all noise sources combined, the likelihood method offers a 12-18% improvement in the residual scatter and maximum deviations from the simulated ground truth.

- Meaningful comparisons between the simulated and on-sky data are difficult to obtain due to their inherent differences, however, they do exhibit similar noise characteristics. For example, the scatter in the residuals between the two methods is reduced primarily as a function of the signal level. The on-sky data do display a ~3 times larger residual scatter between methods for the All noise case, indicating that the differences shown here may be a conservative lower limit, and that further investigation into other detector effects and their interactions with the jump detection and ramp fitting process may be warranted.

- One known caveat to the likelihood method improvements is that our simulated data do not include any cosmic rays, or spatially structured noise such as snowballs or showers. These may have different influences on the different jump detection methods, and preliminary analyses (C. Willott, priv. comm.) indicate that some of the more advanced default jump detection routines present advantages over the likelihood method. This is being actively investigated and potential solutions, such as combining both jump detection algorithms, are under consideration (D. Law, priv. comm.).

- A second caveat to the likelihood method is that it can only operate on datasets with at least 4 groups per integration. With only 3 groups there are only two differenced frames, and individual jumps cannot be robustly resolved and excluded. While this may exclude some science datasets of targets close to the saturation limit, there is potential to broaden the applicability of this method through development of less saturation sensitive subarray readouts (i.e. implementations of MULTISTRIPE, see example for NIRCam https://jwst-docs.stsci.edu/jwst-near-infrared-camera/nircam-instrumentation/nircam-detector-overview/nircam-detector-subarrays/nircam-

multistripe-subarrays).

- For a simulated case with only white noise, we observe a preferential bias towards deeper transit depths that strengthens at lower pixel illuminations for both the default and likelihood methods. This flux-correlated inverse dilution effect has a maximum amplitude of ~150 ppm across our suite of simulations, and has the potential to mimic broad atmospheric absorption features in addition to biasing inferred model measurements.

- Irrespective of jump detection and ramp fitting method, our results agree with existing community knowledge that $1/f$ noise introduces significant correlated noise in transit spectra that can mimic atmospheric features. Importantly, by isolating the $1/f$ noise we observe a differential bias in the 2D spectra at different levels of pixel illumination, and the bias within the spectral trace is visually distinct from the bias outside of the trace. This effect is likely applied during the ramp fitting and attempts to model the $1/f$ noise at the integration-level using unilluminated regions of the detector will not produce an accurate correction.

- Across the broader time-series observation community, group-level $1/f$ corrections are commonplace and may serve to mitigate the observed biases, and reduce the improvements provided by the likelihood methods. Further work should be performed to extend the analysis shown here to a group-level $1/f$ correction. Methods to perform such a correction are available in the current JWST pipeline, but are typically not turned on as a default. As such, testing and integrating these corrections for NIRISS/SOSS should be a key priority to ensure products later than Stage 0 will be actively used by the community.

- While this study focuses on NIRISS SOSS, NIRISS shares a similar detector architecture to NIRCam and NIRSpec. As such, NIRCam/Grism Time-Series and NIRSpec/Bright Object Time-Series observations will likely observe similar benefits to adopting the likelihood method as a standard. Additionally, the effects of $1/f$ noise are likely to be similarly significant, and the development and testing of effective group-level $1/f$ corrections may be necessary. While MIRI has a different detector architecture, and $1/f$ noise is much less significant, there may still be benefits to using the likelihood method.

- Finally, for the simulated case with all noise sources, we find that distribution of measured transit depths is distinctly non-Gaussian. Some of these effects may be mitigated by a group-level $1/f$ correction, but in lieu of further analysis, observers should exercise care when considering such deviations as evidence of difficult to measure atmospheric features across multiple observations or targets.